\documentclass[12pt,article,nofootinbib]{revtex4}
\usepackage{pdfpages}
\usepackage{epsf}
\usepackage{subfigure}
\usepackage{amsmath}
\usepackage{eqnarray,amsmath}
\usepackage{epstopdf}
\usepackage{mathrsfs}
\usepackage{natbib}
\usepackage{amssymb,latexsym}
\usepackage{dcolumn}
\usepackage{graphicx}
\usepackage{color} 
\usepackage{multirow}

\makeatletter
\newcommand*{\rom}[1]{\expandafter\@slowromancap\romannumeral #1@}
\makeatother
\def\be{\begin{equation}}
\def\ee{\end{equation}}
\def\ba{\begin{eqnarray}}
\def\ea{\end{eqnarray}}

\graphicspath{{images/}}

\begin{document}

	\title{\large \bf A new constraint on the Hawking evaporation of primordial black holes in the radiation-dominated era}
	
	\author{Seyed Sajad Tabasi}
    \affiliation{Department of Physics, K.N. Toosi University of Technology, P.O. Box 15875-4416, Tehran, Iran}
	\email{sstabasi98@gmail.com}
	
	\author{Javad T. Firouzjaee}
    \affiliation{Department of Physics, K.N. Toosi University of Technology, P.O. Box 15875-4416, Tehran, Iran}
    \affiliation{PDAT Laboratory, Department of Physics, K. N. Toosi University of Technology, P.O. Box 15875-4416, Tehran, Iran}
    \affiliation{School of Physics, Institute for Research in Fundamental Sciences (IPM), P.O. Box 19395-5531, Tehran, Iran}
	
	\email{firouzjaee@kntu.ac.ir}

\begin{abstract}

 In this paper, we revisit the evaporation and accretion of primordial black holes (PBHs) during cosmic history and compare them to see if both of these processes are constantly active for PBHs or not. Our calculations indicate that during the radiation-dominated era, PBHs absorb ambient radiation due to accretion, and their apparent horizon grows rapidly. This growth causes the Hawking radiation process to practically fail and all the particles that escape as radiation from PBHs to fall back into them. Nevertheless, our emphasis is that the accretion efficiency factor also plays a very important role here and its exact determination is essential. We have shown that the lower mass limit for PBHs that have not yet evaporated should approximately be $10^{14}g$ rather than $10^{15}g$. Finally, we study the effects of Hawking radiation quiescence in cosmology and reject models based on the evaporation of PBHs in the radiation-dominated era.
 
\end{abstract}

\maketitle

\section{Introduction}

Six years ago the decades-long efforts of the LIGO-VIRGO collaboration have taken consequence and led to the first straightforward detection \cite{LIGO2015} of a signal which was for an inspiraling massive black hole binary. This detection and next ones came somewhat as a surprise that most of these binaries were made of fairly massive $30 M_{\odot}$ black holes. Since the star formation scenarios do not necessarily predict mass scale for merging black holes, the theoretical community thought over the question of whether these black holes could be primordial and constitute dark matter. This led people to an interesting conjecture that the LIGO detectors have detected primordial black holes (PBHs).

There are variety of mechanisms for PBHs formation (see e.g. \cite{5,Carr:2020gox} for review). As sets of models focus on inflationary perturbations as a source of PBHs, there are other formation mechanisms \cite{inflation-pbh}, such as cosmic string collapse \cite{cosmic-string}, bubble collisions, domain wall collapse \cite{5,buble} as well as scalar field fragmentation \cite{8,20,23} which can produce plentiful populations of PBHs.
The consequent PBHs can span many orders of magnitude in mass depending on the formation time. People usually apply an important constraint to the PBHs mass range that from hose formed with mass above the Hawking evaporation limit of $10^{15} g$ survive until present. Since this constraint has an important side effect on Diffuse gamma-ray background, Cosmic rays, Neutrinos, Hadron injection, Photodissociation of deuterium, CMB distortion, Photodissociation of deuterium, Present-day relic density \cite{pbh-thesis}, and Big Bang Nucleosynthesis \cite{pbh-from HR}, revisiting their Hawking radiation process is crucial.

After Hawking known paper \cite{hawking-75} on the black hole quantum radiation, different methods for deriving the radiation properties were presented. One can classify these methods in two sets; First, like the Hawking first paper, people compare the quantum vacuum before black hole formation, $|0_{B}>$, and after black hole formation in future infinity, $|0_{A}>$, and they read the radiation properties from
particle creation number expectation value, $<0_{B}|N_k|0_{B}>=\sum_{j}\beta_{jk}$, where $\beta_{jk}$ is the Bogoliubov coefficient which relates the incoming mode from past vacuum to the outgoing mode of the future vacuum.

This quantum field theory approach to black hole radiation, which applies usually to late-time stationary
black holes, is not a suitable method for calculating the Hawking temperature in the case of a fully
dynamical black hole, where one has to solve the field equations in a changing background. The second sets of methods are alternative approaches allowing one to calculate Hawking radiation in a Schwarzschild spacetime without using the field equations, such as finding the related vacuum via the tunneling method \cite{tunneling}. These methods can be extended to calculating quantum fields in the dynamical case \cite{Firouzjaee:2011hi}. Let's look at the essential condition for Hawking radiation in these methods. The main point to have Hawking radiation is to have geometric optic approximation or WKB approximation for and Hawking radiation near the Horizon. For Schwartzhild black hole this condition is satisfied near the apparent or event horizon $R=2GM/c^2$. But in the dynamical black hole can be written in some form eikonal approximation \cite{Visser:2001kq}. This condition can be minimized in terms of adiabatic condition which says as long as one has an approximately exponential relation between the affine parameters on the null generators of past and future null infinity, then subject to a suitable adiabatic condition being satisfied, a Planck-distributed flux of Hawking-like radiation will occur \cite{Barcelo:2010xk}.

Since primordial black hole forms in the FLRW background, they are types of cosmological black holes \cite{Carr:1974nx,Firouzjaee:2010whc}. The black hole properties of a cosmological black hole like PBHs are attributed to the apparent horizon rather than the event horizon \cite{Firouzjaee:2011dn}. Defining the quantum vacuum in the cosmological background which does not have Poincare symmetric or any special killing horizon is not possible or in some special cases is hard \cite{Firouzjaee:2015bqa}. Hence, the tunneling method is appropriate for calculating the Hawking radiation for PBHs \cite{Firouzjaee:2011hi,Firouzjaee:2014zfa}. Nevertheless, there must be WKB or geometric optic approximation as an essential condition for black hole radiation. This condition puts limits on PBHs that can have quantum radiations \cite{Firouzjaee:2015wps}.

Since the issue of the PBHs Hawking radiation has been raised for decades and many works have been done to study the PBHs dynamics and its mass fraction rate based on this radiation, it's necessary to revisit their dynamics in the radiation-dominated era by concerning its accretion flux. We will see our calculations indicate that PBHs accretion put a serious constraint on the PBHs Hawking evaporation.

This paper is organized as follows: first, we review the Hawking radiation and Bondi model accretion concepts and main equations in section II. Then, in section III, the basic subtleties in the black hole radiation process are discussed. Section IV is devoted to comparing the rate of decrease and increase of PBHs mass through these two processes in the cosmological context. Afterward, in section V, the possible effects of accretion flux on previous PBHs models which consider the Hawking radiation were investigated.   
The conclusion and discussions are given in Section VI.

\section{General Equations}

Several models suggest PBHs creation in the early universe. We assume that the formation of PBHs is separated from the expansion of the universe due to the time scale of their creation that is too much shorter than the Hubble time. The PBHs mass can be obtained by the following equation \cite{Carr:2019}

\begin{equation}
\label{e1}
M_{PBH} \sim \frac{c^3t}{G}
\end{equation}
where t is the time after the Big Bang, c is the speed of light, $c\simeq3\times10^{8}m/s$, and G is the gravitational constant, $G\simeq6.67\times10^{-11}m^{3}/kgs^2$.

The most well-known processes that can change the mass of a black hole are evaporation, accretion, and merging. The relevant parameters of a black hole, such as radius, temperature, etc., change with the gain and loss of mass. Immediately after their formation, they start evaporating by the Hawking radiation process. This process causes a black hole to lose its mass by emitting radiation. Also, a black hole, due to its gravitational attraction, can swallow what is around it, thereby increasing its mass. This process, known as accretion, can also lead to binary merging, however, its efficiency is controversial so that we will not consider it in our paper for simplifying.

\subsection{Evaporation}

It is known  that the point mass black holes like those are described with Schwarzschild  and Kerr metric can have quantum radiation which is called Hawking radiation. PBHs which is modeled with these types of metrics evaporate through the Hawking process and lose their mass. The rate of mass change in this way depends on the mass of PBHs. The lower the mass, the higher the radiation emission rate and vice versa. The rate of losing mass for PBHs is given by \cite{Carr:2010}
\begin{equation}
\label{e2}
\frac{dM_{PBH}}{dt}=-f_{eva}4\pi R_{PBH}^{2}c\rho_{PBH},
\end{equation}
where $f_{eva}$ is the evaporation efficiency factor, $R_{PBH}$ is the PBH radius, and $\rho_{PBH}$ can be obtained by
\begin{equation}
\label{e3}
\rho_{PBH}=\frac{\pi^2g_{*}(T_{PBH})k_{B}T_{PBH}^4}{120\hbar^3c^5},
\end{equation}
where g is the number of degrees of freedom, $g_{*}(T)=\sum_{B}g_{B}+7/8\sum_{F}g_{F}$, $k_{B}$ is the Boltzmann constant, $k_{B}\simeq8.62\times10^{-5}eV/K$, $\hbar$ is the reduced Planck constant, $\hbar\simeq6.58\times10^{-15}eV.s$, and $T_{PBH}$ is the PBHs temperature. The emitted particles temperature is exactly the PBH temperature that is given by
\begin{equation}
\label{e4}
T_{PBH}=\frac{\hbar c^3}{8\pi Gk_{B}M_{PBH}}.
\end{equation}

We can rewrite Eq. \eqref{e2} in terms of the temperature and the mass of PBHs
\begin{equation}
\label{e5}
\frac{dM_{PBH}}{dt}=-\frac{f_{eva}2\pi^3g_{*}(T_{PBH})M_{PBH}^2(k_{B}T_{PBH})^4}{15\hbar c^6M_{Pl}^4},
\end{equation}
where $M_{Pl}$ is the Planck mass, $M_{Pl}=1.22\times10^{19}GeV$. If we consider only evaporation as a dominant process that rules PBHs, we can simply calculate their lifetime \cite{Broda:2017}
\begin{equation}
\label{e6}
\tau_{PBH}=\frac{5120\pi G^2M_{PBH}^3}{\hbar c^4}.
\end{equation}

\subsection{Accretion}

As we stated before, we only focus on the accretion process, regardless of merging PBHs, as a process that causes PBHs to gain mass. By assuming that all PBHs have been formed in the radiation-dominated era, their mass increases by falling the photons from the thermal bath into them. The spherically symmetrical accretion rate that has been derived by Bondi \cite{Bondi}  is
\begin{equation}
\label{e7}
\frac{dM_{PBH}}{dt}=f_{acc}4\pi R_{PBH}^{2}c\rho_{R},
\end{equation}
where $f_{acc}$ is the accretion efficiency factor and $\rho_{U}$ is the radiation mass density at a high temperature which approximately is
\begin{equation}
\label{e8}
\rho_{R}=\frac{\pi^2g_{*}k_{B}^4T_{U}^4}{30\hbar^3c^5}
\end{equation}
where $T_{U}$ is the thermal bath temperature.
We write Eq. \eqref{e7} in terms of PBHs mass and temperature, so we have
\begin{equation}
\label{e9}
\frac{dM_{PBH}}{dt}=\frac{f_{acc}8\pi^3g_{*}(T_{U})M_{PBH}^2(k_{B}T_{U})^4}{15\hbar c^6M_{Pl}^4}.
\end{equation}

The crucial question that pops up here is that do both accretion and evaporation always play an important role in PBHs dynamics? In the following, we will use two perspectives to answer this question, one is the quantum perspective and the other is the cosmological perspective. The quantum standpoint that we explain is based on the previous works like  \cite{Firouzjaee:2014zfa, Firouzjaee:2015bqa} that elaborate the dynamical horizon of Schwarzschild black holes. Nevertheless, in the cosmological standpoint section, we elaborate its cosmological consequences in the radiation-dominated era.

\section{Quantum Standpoint}

The first derivation for the black hole radiation by Hawking is based on assuming that a future eternal event horizon forms, and that the subsequent exterior geometry is static.
Developing the semi-classical physics in curved spacetime helped us to calculate the renormalized energy-momentum tensor, $<T_{\mu \nu}>$, and extract the spectrum of the Hawking radiation \cite{book,Firouzjaee:2015bqa}. To know the origin of the Hawking radiation, extrapolating the $<T_{\mu \nu}>$ properties in the spherically symmetric case proposed that the Hawking effect is associated with the apparent horizon rather than the event horizon \cite{Hajicek:1986hn}. 

Extending the Hawking radiation process by tunneling method \cite{tunneling} opened our hand to derive the black hole quantum radiation for non-stationary cases. We know that the quantum fluctuations around the horizon are of the order of the Planck length \cite{Jacobson:1993}. To have real particles from virtual particles in these quantum fluctuations, the tunneling scenario proposes a known process around the black holes \cite{tunneling}. Tunneling calculation is based on the WKB approximation assumed for the light coming from the apparent horizon to the observer \cite{Visser:2001kq}. The required condition for black hole radiation can be minimized in terms of the adiabatic condition which says as long as one has an approximately exponential relation between the affine parameters on the null generators of past and future null infinity, then subject to a suitable adiabatic condition being satisfied, a Planck-distributed flux of Hawking-like radiation will occur \cite{Barcelo:2010xk}.

All these methods give the tunneling probability rate $\Gamma_{em}\sim e^{-\frac{\hbar w}{k_B T_H}}$ turns to be a thermal form where $\hbar w$ is the emitted particle energy and $T_H$ is the radiation temperature which is related to the black hole's surface gravity$T_h=\frac{\kappa}{2 \pi}$. The basic point is that if a black hole is in the dynamical phase for example due to the accretion \cite{Ashtekar:2003hk}, one cannot have the essential conditions such as WKB approximation our adiabatic condition around the apparent horizon for Hawking radiation. This leads to quenching the black hole radiation in the dynamical phase \cite{Firouzjaee:2014zfa,Firouzjaee:2015wps}. 

\begin{figure}[h]
	\begin{center}
		\includegraphics[scale = 0.9]{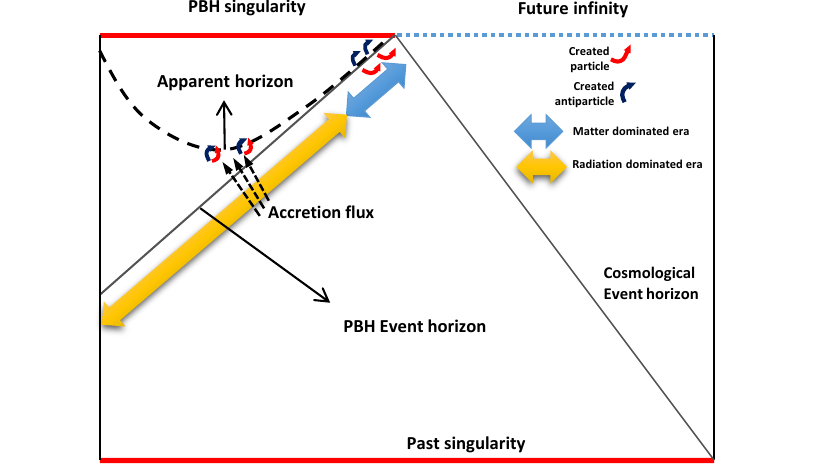}
		\hspace*{10mm} \caption{ \label{tpenrose}
			Penrose diagram for a primordial black hole in a $\Lambda CDM$ expanding universe, where there is some matter and radiation accretion flux, but it dies away to zero in the far future (matter-dominated era). Note that we do not represent backreaction effects in this picture. Thus, the black hole dynamical horizon tends to be event horizon in the matter-dominated era.}
	\end{center}
\end{figure}

Moreover, in any dynamical spacetime with the in-falling matter, the apparent (dynamical) horizon is inside the event horizon \cite{Ashtekar:2003hk}, thus any virtual pair particle created due to these effects cannot fall outside the event horizon, become real, and reach future infinity. You can see this feature in Fig. (\ref{tpenrose}) for PBHs. This figure is the Penrose diagram for a PBH which is formed in the radiation-dominated era. PBHs' apparent horizon grows in the radiation-dominated era till its accretion flux is stopped by the end of this era. After that in the matter-dominated era, these black holes start to radiate.

\section{Cosmological Standpoint}

According to Eq. \eqref{e5} and Eq. \eqref{e9}, it is obvious that the only difference that determines which process is dominant is the thermal bath and PBHs temperature \cite{Masina:2020}. During the radiation-dominated era, the expansion time of the universe in terms of the temperature of the universe is given by \cite{Manolis:2002}

\begin{equation}
\label{e10}
T_{U}(Kelvin)\simeq\frac{1.52\times 10^{10}}{\sqrt{t(sec)}}.
\end{equation}

By considering that all PBHs are Schwarzschild black holes, by using \eqref{e1} and Eq. \eqref{e4} we have the PBHs temperature in terms of the time after the Big Bang

\begin{equation}
\label{e11}
T_{PBH}(Kelvin)\simeq\frac{3.04\times 10^{-13}}{t(sec)}.
\end{equation}

We should compare the two temperatures we got. In the time that the universe temperature is larger than the PBHs temperature, the accretion process dominates over the evaporation process. Then we can conclude that PBHs gain mass and their radius increase respectively. On the other hand, in the time when the PBHs temperature is larger than the thermal bath, PBHs lose their mass and their radius gets shorter.  With a simple calculation, we can find the time of equality $(T_{PBH}=T_{U})$, $t_{eq} \sim 10^{-44}s$. This time is much earlier than the time of the inflation. PBHs that are formed before the end of the inflation are not important for us because of their negligible density \cite{Pablo:2021}. We can see that from the beginning of the radiation-dominated era until the start of the matter-dominated era, the universe temperature is much bigger than the PBHs temperature. Hence, the accretion process is the process that rules the mass evolution of PBHs during the radiation-dominated era. During the matter-dominated era, people have shown that accreting the environment's radiation does not have any significant effect on the mass of PBHs \cite{Maxim:2020}. 

We know that there are only three possible mass windows for PBHs with a mass larger than $10^{15}g$: (i) $10^{16}-10^{17}g$, (ii) $10^{20}-10^{24}g$, and (iii) $10^{33}-10^{36}g$ for explaining the whole dark matter \cite{Carr:2016}. The other ranges are eliminated by various constraints that are based on theoretical and observational works. In Table (\rom{1}), we compare the PBHs temperature of upper and lower bound mass windows with the universe temperature at the time of their formation. 

\begin{table}[h!]
	\begin{center}
		\begin{tabular}{ | m{1.5cm} | m{1.9cm}| m{1.9cm} | m{1.9cm} | m{1.9cm} |  m{1.9cm} | m{1.9cm} |}
			\hline
			\small$M_{i}(g)$ &\small $10^{16}$ &\small $10^{17}$ &\small $10^{20}$ &\small $10^{24}$ &\small $10^{33}$ &\small $10^{36}$ \\
			\hline
			\small$t_{i}(s)$ & \footnotesize$2.47\times10^{-23}$ &\footnotesize $2.47\times10^{-22}$ &\footnotesize $2.47\times10^{-19}$ &\footnotesize $2.47\times10^{-15}$ &\footnotesize $2.47\times10^{-6}$ &\footnotesize $2.47\times10^{-3}$\\
			\hline
			\small$T_{PBH}(K)$  & \footnotesize$1.23\times10^{10}$  &\footnotesize $1.23\times10^{9}$&\footnotesize $1.23\times10^{6}$&\footnotesize $1.23\times10^{2}$&\footnotesize $1.23\times10^{-7}$&\footnotesize $1.23\times10^{-10}$ \\
			\hline
			\small$T_{U}(K)$  &\footnotesize $3.06\times10^{21}$  &\footnotesize $9.67\times10^{20}$  &\footnotesize$3.06\times10^{19}$  &\footnotesize$3.06\times10^{17}$  &\footnotesize$9.67\times10^{12}$  &\footnotesize$9.67\times10^{11}$  \\
			\hline
			
		\end{tabular}
		\label{t1}
		\caption{As shown in this table, in all mass windows of PBHs to provide the dark matter, the temperature of the universe is much higher than the temperature of PBHs, therefore only the accretion process can be effective. }
		\label{table:1}
	\end{center}
\end{table}

Now, it should be derived that for the accretion and mass increase of PBHs, how much their radius changes. This calculation is important because it determines whether the accretion strength is enough to prevent Hawking radiation during the radiation-dominated era. We do these calculations in two ways. Once we use the model used in \cite{Nayak:2009wk} for the radiation-dominated era and once again we directly enter the effects of the growth of the universe through the way that has been used in \cite{Nayak:2011sk}. 

\subsection{Model \rom{1}}
By using Eq. \eqref{e7} and substituting $\rho_r$ and $R_{PBH}$ with their well-defined equations in the standard cosmology, $\rho=3\Dot{a}^2/8\pi Ga^2$ and $R_{PBH}=2GM_{PBH}/c^2$, we can have the accretion rate in terms of the mass of a PBH and the time after the Big Bang

\begin{equation}
\label{e12}
\frac{dM_{PBH}}{dt}=\frac{3f_{acc}G}{2c^3}\frac{M_{PBH}^2}{t^2}.
\end{equation}
By integrating from Eq. \eqref{e12} we have
\begin{equation}
\label{e13}
M(t)=(M_{i}^{-1}+\frac{3f_{acc}G}{2c^3}(\frac{1}{t}-\frac{1}{t_{i}}))^{-1}.
\end{equation}
Then we substitute Eq. \eqref{e1} into the Eq. \eqref{e13} and we get

\begin{equation}
\label{e14}
M(t)=M_{i}(1+\frac{3f_{acc}}{2}(\frac{t_{i}}{t}-1))^{-1}.
\end{equation}

Eq. \eqref{e14} simply gives us the final mass of PBHs with the initial mass $M_{i}$ at time $t$ after the Big Bang. Since we are looking at the evaluation of PBHs during the radiation-dominated era, the value of $t_{1}$ is the end of the radiation-dominated era, $t_{1}\simeq1.48\times10^{12}s$. Moreover, for the accretion efficiency factor, we should determine it with a value lower than 2/3.

At first glance, it seems strange that the ratio of final mass to initial mass per each accretion efficiency factor is the same for different initial masses. It should be noted that the final time is fixed, and on the other hand, the initial mass is directly related to the initial time, therefore, by determining the initial mass, we also set the initial time.

It is worth mentioning that the efficiency of disk accretion is significantly more than spherical accretion\cite{Kamionkovski,Shapiro} that we have considered in our calculations. Ultimately, $f_{acc}$ is the important parameter that can determine the final mass of PBHs at the end of the radiation-dominated era in Bondi accretion. The larger the $f_{acc}$, the greater the final mass. We know that the growth of the mass is proportional to the growth of the Schwarzschild radius. Thus, for PBHs radius we have

\begin{equation}
\label{e15}
R(t)=R_{i}(1+\frac{3f_{acc}}{2}(\frac{t_{i}}{t_{f}}-1))^{-1}.
\end{equation}

This means that if, for example, the mass of a PBH increases 10 times by the end of the radiation-dominated era, its radius will increase by the same factor. Once again, we should mention that taking different accretion efficiency factors can change the result significantly. One of the things that we want to emphasize in this paper is that people cannot be careless about the value of $f_{acc}$.

\subsection{Model \rom{2}}
Furthermore, we examine another model and compare its results with the model mentioned in the Eq.\eqref{e12}. Considering spherically symmetrical accretion, we use the Eq.\eqref{e7}, and since we focus on the accretion of radiation, we consider the value of $v={c}/{\sqrt{3}}$ and $R_{PBH}=R_{s}$. Thus we can rewrite the last equation as

\begin{figure*}[ht]
 \label{fig:4}
\subfigure[]{
  \includegraphics[width=75mm]{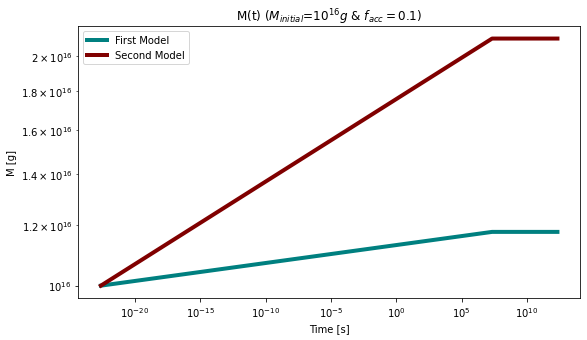}
}
\hspace{0mm}
\subfigure[]{
  \includegraphics[width=75mm]{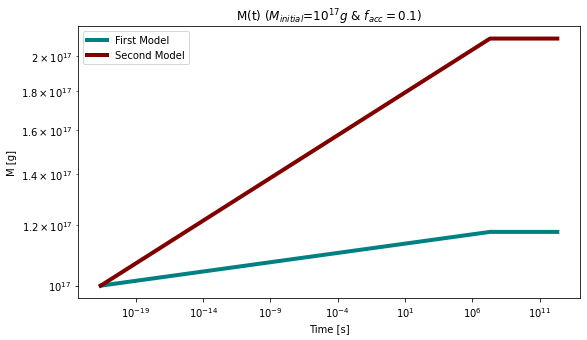}
}
\hspace{0mm}
\subfigure[]{
  \includegraphics[width=75mm]{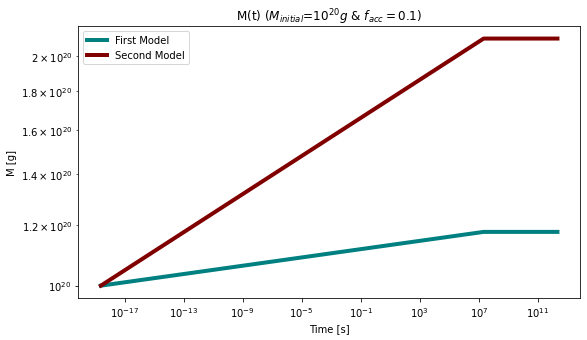}
}
\subfigure[]{
  \includegraphics[width=75mm]{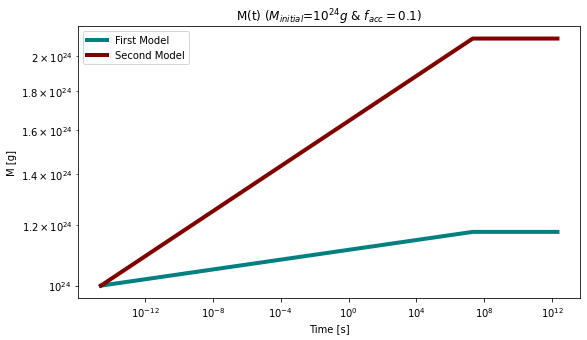}
}
\subfigure[]{
  \includegraphics[width=75mm]{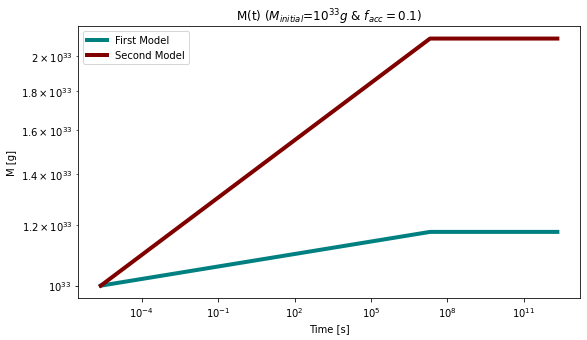}
}
\subfigure[]{
  \includegraphics[width=75mm]{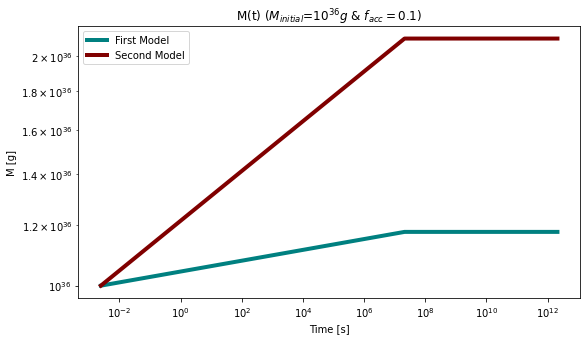}
}
\begin{center}
\caption{\begin{footnotesize} This figure points out the competition accretion between two mentioned models in the radiation-dominated era for six initial masses (a)$10^{16}g$, (b)$10^{17}g$, (c)$10^{20}g$, (d)$10^{24}g$,(e)$10^{33}g$, and (f)$10^{36}g$. The second model obviously is more effective than the first model. In all figures the accretion efficiency factor has been considered 0.1.\end{footnotesize}}
\end{center}
\end{figure*}

\begin{equation}
\label{e16}
\frac{dM_{PBH}}{dt}=16\pi G^2 M_{PBH}^2\rho_{r} (\frac{c}{\sqrt{3}})^{-3} f_{acc}.
\end{equation}

For a more effective comparison of the two models, it is necessary to calculate $M(t)$ in the second model too. By using Friedmann equations, we have $\rho\propto{a}^{-3(1+\omega)}$ that $a$ is the scale factor and $\omega$ is the equation of state parameter which is equal to 1/3 for radiation. On the other hand, we know for a flat universe $a\propto t^{2/3(1+\omega)}$, so Eq.\eqref{e16} can be written as follows \cite{Nayak:2011sk}

\begin{equation}
\label{e17}
\frac{dM_{PBH}}{dt}=16\pi G^2 \rho_{cr} \Omega_{r}^0 (\frac{c}{\sqrt{3}})^{-3} f_{acc}(t_{1}^{-\frac{2}{3}} t_{2}^\frac{8}{3} e^{-4H_{0}(t_{2}-t_{0})}) (\frac{M_{PBH}}{t})^2,
\end{equation}

where $\rho_{cr}=9.2\times10^{-30}{g}/{cm^3}$ is the critical energy density, $\Omega_{r}^0=4.2\times10^{-5}$ is relative contribution of relativistic particles, $H_{0}\simeq70 s^{-1} Mpc^{-1}$ is Hubble parameter, $t_{2}=2.4\times10^{17}s$ is the time of end of the matter-dominated era, and $t_{0}=4.4\times10^{17}s$ is the present time \cite{Rubakov:2017xzr}. Eventually, the final mass of PBHs in the radiation-dominated era is calculated by solving the differential equation of Eq.\eqref{e17}

\begin{equation}
\label{e18}
M(t)=M_{i}(1+5.28f_{acc}(\frac{t_{i}}{t}-1))^{-1}.
\end{equation}

As a result, we can calculate $R_{t}$ for Model \rom{2} as follows

\begin{equation}
\label{e19}
R(t)=R_{i}(1+5.28f_{acc}(\frac{t_{i}}{t}-1))^{-1}.
\end{equation}

Now that the equations of both models are obtained, the best thing to do is to compare the mass increase of these two models with each other during the radiation-dominated era. Figure (II) demonstrates well how PBHs evolve in terms of mass for each model. It affirms that PBHs masses in Model (II) increase more against Model (I). Table (II) shows the final radii of PBHs that their initial masses have been considered in Table (I) at the end of the radiation-dominated era with three different accretion efficiency factors 0.05, 0.1, and 0.15.

\begin{table}[h!]
	\begin{center}
		\begin{tabular}{ | m{1.5cm} | m{2.1cm}| m{1.3cm} | m{1.9cm} | m{1.9cm} |  m{1.9cm} |}
			\hline
			\centering
			\multirow{2}{*}{\small$M_{i}(g)$} &\centering \multirow{2}{*}{\small $R_{i}(m)$} &\multirow{2}{*}{\small $Model$ }& \multicolumn{3}{|c|}{\small $Final Raduis$} \\
			\cline{4-6}
			& & &\small $f_{acc}=0.05$ &\small $f_{acc}=0.1$ &\small $f_{acc}=0.15$\\
			\hline
			\centering
			\multirow{2}{*}{$10^{16}$} & \multirow{2}{*}{\small $1.48\times10^{-11}$} &\footnotesize \centering$ \rom{1}$ &\footnotesize $1.60\times 10^{-11}$ &\footnotesize $1.74\times10^{-11}$ &\footnotesize $1.91\times10^{-11}$ \\
			\cline{3-6}
			& &\footnotesize\centering $\rom{2}$ &\footnotesize $2.01\times10^{-11}$ &\footnotesize $3.14\times10^{-11}$ &\footnotesize $7.14\times10^{-11}$\\
			\hline
			\centering
			\multirow{2}{*}{$10^{17}$} & \multirow{2}{*}{\small $1.48\times10^{-10}$} &\footnotesize \centering $\rom{1}$ &\footnotesize $1.60\times 10^{-10}$ &\footnotesize $1.74\times10^{-10}$ &\footnotesize $1.91\times10^{-10}$ \\
			\cline{3-6}
			& &\footnotesize \centering$\rom{2}$ &\footnotesize $2.01\times10^{-10}$ &\footnotesize $3.14\times10^{-10}$ &\footnotesize $7.14\times10^{-10}$\\
			\hline
			\centering
			\multirow{2}{*}{$10^{20}$} & \multirow{2}{*}{\small $1.48\times10^{-17}$} &\footnotesize \centering$\rom{1}$ &\footnotesize $1.60\times 10^{-7}$ &\footnotesize $1.74\times10^{-7}$ &\footnotesize $1.91\times10^{-7}$ \\
			\cline{3-6}
			& &\footnotesize \centering$\rom{2}$ &\footnotesize $2.01\times10^{-7}$ &\footnotesize $3.14\times10^{-7}$ &\footnotesize $7.14\times10^{-7}$\\
			\hline
			\centering
			\multirow{2}{*}{$10^{24}$} & \multirow{2}{*}{\small $1.48\times10^{-3}$} &\footnotesize \centering$\rom{1}$ &\footnotesize $1.60\times 10^{-3}$ &\footnotesize $1.74\times10^{-3}$ &\footnotesize $1.91\times10^{-3}$ \\
			\cline{3-6}
			& &\footnotesize \centering$\rom{2}$ &\footnotesize $2.01\times10^{-3}$ &\footnotesize $3.14\times10^{-3}$ &\footnotesize $7.14\times10^{-3}$\\
			\hline
			\centering
			\multirow{2}{*}{$10^{33}$} & \multirow{2}{*}{\small $1.48\times10^{6}$} &\footnotesize \centering$\rom{1}$ &\footnotesize $1.60\times 10^{6}$ &\footnotesize $1.74\times10^{6}$ &\footnotesize $1.91\times10^{6}$ \\
			\cline{3-6}
			& &\footnotesize \centering$\rom{2}$ &\footnotesize $2.01\times10^{6}$ &\footnotesize $3.14\times10^{6}$ &\footnotesize $7.14\times10^{6}$\\
			\hline
			\centering
			\multirow{2}{*}{$10^{36}$} & \multirow{2}{*}{\small $1.48\times10^{9}$} &\footnotesize \centering$\rom{1}$ &\footnotesize $1.60\times 10^{9}$ &\footnotesize $1.47\times10^{9}$ &\footnotesize $1.91\times10^{9}$ \\
			\cline{3-6}
			& &\footnotesize \centering $\rom{2}$ &\footnotesize $2.01\times10^{9}$ &\footnotesize $3.14\times10^{9}$ &\footnotesize $7.14\times10^{9}$\\
			\hline
		\end{tabular}
		\label{t2}
		\caption{This table illustrates final raduis at the end of radiation-dominated for two mentioned models by considering different accretion efficiency factors.}
		\label{table:1}
	\end{center}
\end{table}

The growth of the PBHs radius causes a huge problem for the PBHs evaporation process. As discussed in the previous section, we write the Hawking radiation equations for an adiabatic horizon. Many papers tried to use this phenomenon for PBHs, but they missed something named accretion.

By the accretion of PBHs, they gain mass and their radius grows permanently during the radiation-dominated era. Therefore, PBHs in the radiation-dominated era do not have the adiabatic condition on the (apparent) horizon and hence we cannot write the regular Hawking radiation formulas for them. Therefore, if the growth rate of the (apparent) horizon is huge that at each time interval the radius of PBHs grows much more than the order of Planck length, the produced escaping particles fall back into PBHs and Hawking radiation turns off. This situation remains until the end of the radiation-dominated era. Since we have shown in Table (II) that the mass and radius growth rates of PBHs are extremely dependent on the accretion efficiency factor, it can be concluded that the larger the $f_{acc}$, the lower the probability of occurring tunneling and consequently evaporation. Finally, the importance of exactly determining the $f_{acc}$ has been cleared because it is practically what determines which process rules PBHs, accretion, evaporation, or both of them.

\section{Possible Effects}

So far we have shown that accretion is the dominant process of PBHs during the radiation-dominated era and the final mass of a PBH at the end of the radiation-dominated era depends on the accretion efficiency factor. This has a serious impact on two issues. One is the effect on the lower mass limit of non-evaporated PBHs, and the second is the effect on the models that suggest the production of particles from the evaporation of PBHs during the radiation-dominated era. In the following, we will investigate each of these two effects separately.

Although neglecting Hawking radiation may seem scary at first, it also raises hope too. It has been emphasized in previous works that PBHs with a mass of less than $10^{15}g$ have completely evaporated and cannot be observed now. Because of that, people almost always consider the evaporation process as an effective process. In this work, we have shown that such an idea is wrong because of the accretion. Given the fact that PBHs do not start to evaporate until the beginning of the matter-dominated era, their lifetime calculations must be considered from the beginning of this epoch. Therefore, the remained possible PBHs mass windows for explaining the dark matter extend. By using Eq. \eqref{e6}, it can be shown that the lower mass limit for PBHs that have not yet evaporated should approximately be $10^{14}g$. In the other words, we should investigate the possibility of the existence of PBHs in the mass range of $10^{14}g$ to $10^{15}g$ too.

The other significant effect is that over the years, there have been various suggestions that the process of the evaporation of PBHs could answer the different questions posed in cosmology. Each of them requires a specific initial mass range for PBHs. In this paper, we have shown that evaporation of PBHs in the radiation-dominated era is off because of the accretion. This refutes all the suggestions that utilize Hawking radiation of PBHs during the radiation-dominated era. Here, we mention some of the famous ones.
\begin{itemize}
    \item CMB distortions are slightly observational deviations from the black body spectrum of the CMB. One of the important suggestions for explaining the CMB distortions is the evaporation of PBHs. As we can see in \cite{Tashiro} and \cite{Mather}, photons emitted from PBHs with the mass $10^{11}g$ to $10^{13}g$ in the redshift between $z\simeq10^{6}$ and $z\simeq10^{3}$ can cause the CMB distortions. 
    \item The dominance of matter over antimatter is still a huge problem in physics. Baryogenesis is a process that causes baryonic asymmetry to provide this dominance. \cite{Hooper} Suggests that evaporation of PBHs with the mass range of $10^{5}g$ to $10^{9}g$ during the radiation-dominated era can elaborate GUT baryogenesis. 
    \item Produced photons from PBHs with the mass $10^{10}g$ to $10^{13}g$ between the end of nucleosynthesis and recombination can photodissociate deuterium. This has been investigated in \cite{Clancy}. They have described it with braneworld cosmology in detail.
    \item Dark matter is a main component of the universe that has not yet been properly explained. One approach to explaining dark matter is to consider the dark matter as special particles that we have not yet been able to detect. Works like \cite{Gondolo} claim that PBHs can produce dark matter particles via Hawking radiation mechanism during the radiation-dominated era and early matter-dominated era.
    \item The study of \cite{Kim} indicates that the diffuse gamma-ray background can relate to  PBHs with the mass $2\times10^{13}g$ to $5\times10^{14}g$ which have been evaporated from $z\simeq700$ until now. The production of the cosmic rays \cite{Gibbon} and the diffuse neutrino background\cite{Bugaev} by evaporation of PBHs circumstances are similar to the diffuse gamma-ray background.  
\end{itemize}

Now, it should be noted that as we proved before, the suggestions that are based on Hawking evaporation of PBHs during the radiation-dominated era are rebutted. Consequently, from the examples we just mentioned, evaporation of PBHs cannot explain CMB distortions, Baryogenesis, photodissociation of deuterium, and dark matter particles produced in the radiation-dominated era. 

One way to save these models is to select a very small accretion efficiency factor so that the accretion rate can be neglected.  However, our work does not affect the suggestions that occur after the end of the radiation-dominated era. This matter should be investigated later though.

Another way is to look for theories for the formation of PBHs in which PBHs are produced whose temperature at the time of formation is about or greater than the cosmic background temperature. In this case, it can be claimed that PBHs will have Hawking radiation. In this case, the accretion cannot turn off the Hawking radiation, and particles escape from the PBHs as Hawking radiation and enter the cosmic environment. Anyhow, it can be said with certainty that PBHs formed according to Eq. \eqref{e1} do not have the ability to evaporate in the radiation-dominated era.

\section{Conclusions}
In this work, we have revisited and revised the two processes that can change the mass of PBHs, accretion and evaporation. Considering Eq. \eqref{e5} and Eq. \eqref{e9}, one can conclude that the only parameter that determines which process rules PBHs is the temperature difference between PBHs and the universe. Comparing the temperature of PBHs and the universe's temperature saw that the temperature of the universe during the radiation-dominated era is much more than the temperature of PBHs. For instance, we have brought up the remained mass windows of PBHs which can explain dark matter in Table (I). Thus, the mass of PBHs permanently increases during the radiation-dominated era and accretion is the dominant process.

Next, the question we tried to answer was whether, with the persistent increase in the mass of PBHs, the evaporation process occurs at all. We have used two different approaches to answer this question. First, we have used the Quantum standpoint and second, we have used the cosmological standpoint. In Figure (I) we have illustrated the growth of the PBHs (apparent) horizon with a Penrose diagram in a $\Lambda$CDM expanding universe. Afterward, we applied the Bondi spherical accretion model to derive the final mass and final radius of PBHs at the end of the radiation-dominated era with two different models. We have demonstrated that the final mass and respectively the final radius totally depends on the accretion efficiency factor. We have considered the remained mass windows of PBHs and their radii for explaining the dark matter in Table (II) again and have elaborated on the dependency. The growth of the (apparent) horizon by accretion turns off PBHs evaporation and the escaping particles fall back into PBHs immediately. Thus, our calculations indicate that PBHs accretion put a serious constraint on the PBHs Hawking evaporation.

In the end, we have investigated the effects of turning off the Hawking evaporation. We illustrated that instead of considering PBHs with larger than $10^{15}g$ initial mass as non-evaporated PBHs, we should expand this lower bound to $10^{14}g$ for the reason that PBHs start evaporating from the beginning of the matter-dominated era and they gain mass during the radiation-dominated era. It is important to note that if PBHs are observed in the future, the accretion efficiency factor can be estimated by only measuring their mass. we questioned the models that are based on Hawking evaporation of PBHs during the radiation-dominated era such as CMB distortions, baryogenesis, and dark matter particles production by PBHs due to the Hawking radiation.

This is the beginning of an attempt to re-examine the processes involved in changing the mass of PBHs with a dynamic model of their apparent horizon at different eras in the history of the universe. Specifically in this paper, we used a toy model to change the apparent horizon through time to study the accretion and evaporation of PBHs to see if these processes are always active. In this article, we also emphasized that determining the amount of accretion efficiency factor is very important in the mass growth of PBHs. This paper is the first step to taking a closer look at how the behavior of PBHs over time may require a more careful and simplistic review. In the following works, for the radiation-dominated era, in addition to radiation, we will also consider the effect of matter accretion, and we are also curious to see whether accretion and evaporation are active in the matter-dominated era, by considering both matter and radiation. Moreover, we will show what parameters will play a significant role in this comparison and how important it is to determine their value.

\section*{Acknowledgements}

We are thankful to Mahsa Berahman, PDAT Laboratory, K. N. Toosi University of Technology, for her suggestions which greatly helped in preparing this manuscript.

\end{document}